\documentclass[floats,prd,aps,amssymb,nofootinbib,showkeys]{revtex4}
\usepackage{amssymb}
\usepackage{graphics}
\usepackage{amsmath}
\usepackage{amsfonts}
\usepackage{bm}
\usepackage[]{latexsym}

\DeclareMathOperator{\csch}{csch}

\newcommand{\be}{\begin{equation}}\newcommand{\ee}{\end{equation}}
\newcommand{\bea}{\begin{eqnarray}}\newcommand{\eea}{\end{eqnarray}}
\newcommand{\brr}{\begin{array}}\newcommand{\err}{\end{array}}
\newcommand{\bit}{\begin{itemize}}\newcommand{\eit}{\end{itemize}}
\newcommand{\ben}{\begin{enumerate}}\newcommand{\een}{\end{enumerate}}

\newcommand{\ba}{\begin{array}}
\newcommand{\ea}{\end{array}}

\def\lf{\left}

\def\ri{\right}

\def\si{\sigma}
\def\Om{\Omega}

\def\1{{_{1}}}\def\2{{_{2}}}
\def\bk{{\bf {k}}}\def\bx{{\bf {x}}}

\def\noHe0{:\;\!\!\;\!\!:H_e(0):\;\!\!\;\!\!:}
\def\noHm0{:\;\!\!\;\!\!:H_\mu(0):\;\!\!\;\!\!:}

\def\lf{\left}

\def\ri{\right}

\def\si{\sigma}
\def\Om{\Omega}

\def\1{{_{1}}}\def\2{{_{2}}}

\def\bogubst{{\rho}^{*}_{\bk}}
\def\bogub{{\rho}_{\bk}}
\def\bogvb{{\lambda}_{\bk}}

\def\ebogomeno{e^{i\left(\omega_{\bk,2}-\omega_{\bk,1}\right) t}}
\def\ebogopiu{e^{i\left(\omega_{k,1}+\omega_{k,2}\right) t}}
\def\uuu{U}

\def\lf{\left}

\def\ri{\right}

\def\si{\sigma}
\def\Om{\Omega}

\def\1{{_{1}}}\def\2{{_{2}}}

\def\bogubst{{\rho}^{\bf{k}\, *}_{12}}
\def\bogub{{\rho}^{\bf{k}}_{12}}
\def\bogvb{{\lambda}^{\bf{k}}_{12}}

\def\bogocoeffAlpha{{\cal A}_{(\Om,\Om'),\,\vec{k}}^{(\si,\si')}}

\def\bogocoeffBeta{{\cal B}_{(\Om,\Om'),\,\vec{k}}^{(\si,\si')}}

\def\bogocoeffAlphamensigstarkapdue{{\cal A}_{(\Om,\Om'),\,\vec{k}}^{(\si,\si')}}

\def\bogocoeffBetamensigstarkapdue{{\cal B}_{(\Om,\Om'),\,\vec{k}}^{(\si,\si')}}

\def\uuu{U}

\begin{document}
\title{$q$-generalized Tsallis thermostatistics in Unruh effect for mixed fields}

\author{Giuseppe Gaetano Luciano\footnote{email: gluciano@sa.infn.it (corresponding author)}$^{\hspace{0.1mm}1,2}$ and Massimo Blasone\footnote{email: blasone@sa.infn.it}$^{\hspace{0.1mm}1,2}$} 
\affiliation
{$^1$Dipartimento di Fisica, Universit\`a di Salerno, Via Giovanni Paolo II, 132 I-84084 Fisciano (SA), Italy.\\ 
$^2$INFN, Sezione di Napoli, Gruppo collegato di Salerno, Via Giovanni Paolo II, 132 I-84084 Fisciano (SA), Italy.}

\date{\today}

\begin{abstract}
It was shown that the particle distribution detected
by a uniformly accelerated observer in the inertial vacuum (Unruh effect)
deviates from the pure Planckian spectrum 
when considering the superposition of fields with different masses. 
Here we elaborate on the statistical origin of this phenomenon. 
In a suitable regime, we provide an effective description of
the emergent distribution in terms of the nonextensive $q$-generalized statistics based on Tsallis entropy. This picture allows us to establish a nontrivial relation between the $q$-entropic index and the characteristic mixing parameters $\sin\theta$ and $\Delta m$. In particular, we infer that $q<1$, indicating the superadditive feature of Tsallis entropy in this framework. We discuss our result in connection with the entangled condensate structure acquired by the
quantum vacuum for mixed fields. 
\end{abstract}


\date{\today}

\maketitle

\section{Introduction}
The phenomenon of quantum mixing, i.e. the superposition
of particle states with different masses, is among the most challenging topics in Particle Physics. In the Standard Model, it appears in the quark sector through Kobayashi-Maskawa matrix~\cite{KM}, 
a three-family generalization of Cabibbo mixing matrix 
between $d$ and $s$ quarks~\cite{Cabi}. On the other hand, 
convincing evidences of flavor mixing and oscillations in the neutrino sector have been
provided in recent years by Super-Kamiokande~\cite{SKK} 
and SNO experiments~\cite{SNO}, confirming
Pontecorvo's pioneering idea~\cite{Pont} and opening
a window into physics beyond the Standard Model. 

Recently, the relevance of mixing transformations 
has prompted their study from a more fundamental
field-theoretical (QFT) perspective. QFT effects on flavor mixing have been analyzed both for Dirac fermions~\cite{BV95} and bosons~\cite{BlasCap}. This has uncovered the limits of the
original quantum mechanical approach by pointing out the orthogonality 
between the vacuum for fields with definite flavor and that for fields with definite mass, the former becoming a condensate
of particle-antiparticle pairs.
The properties of flavor vacuum
have been further explored in~\cite{Cabo}, where
it has been shown that the Fock space for flavor fields cannot be obtained 
by the direct product of the Fock spaces for massive
fields. Therefore, the nontrivial nature of mixing
appears as a genuine QFT feature boiling down to the 
nonfactorizability of the flavor states in terms
of those with definite mass, including the vacuum state (flavor vacuum).

All of the above studies have been developed in
Minkowski spacetime. The QFT approach
to mixing has been extended to 
Rindler (uniformly accelerated) metric in~\cite{Luciano,NonTN}
and to curved background in~\cite{Quaranta}.
In particular, in~\cite{Luciano,NonTN} it has been
found that the vacuum condensate detected by the 
Rindler observer due to Unruh effect~\cite{Unruh}
deviates from the Planckian density profile
in the presence of mixed fields, the departure being dependent 
on the mass difference and the mixing angle. Such a result
has been originally interpreted as a breakdown of the thermality of Unruh radiation 
for mixed fields. In passing, we mention that 
unconventional behaviors of Unruh effect are 
not entirely unusual in the literature, see for instance~\cite{Dop,Hammad,deformed1,deformedPet} and possible
implications for particle decays~\cite{Decay1,Decay2,Decay3}.  

In its traditional form, the particle number spectrum 
of Unruh condensate follows 
the rules of Boltzmann-Gibbs
statistics. However, in~\cite{Tsallis1,Tsallis2,Tsallis3,Tsallis4}
it has been argued that systems exhibiting long-range
interactions and/or spacetime entanglement, either on quantum or classical grounds, require a
generalization of Boltzmann-Gibbs theory to the so called 
nonextensive Tsallis $q$-thermostatistics.
This occurs through a suitable
(nonadditive) redefinition of the entropy, which still
recovers Boltzmann-Gibbs formula in the $q=1$ limit.    
The $q$-generalized statistical mechanics proposed
by Tsallis has provided encouraging results in
describing a broad class of complex systems, such as
self-gravitating stellar systems~\cite{App1,App3}, black holes~\cite{Tsallis3}, the cosmic background radiation~\cite{App7,App8}, 
low-dimensional dissipative systems~\cite{Tsallis4}, solar neutrinos~\cite{App11}, polymer chains~\cite{Polch} and modified cosmological models~\cite{App13,App14}, among others. Furthermore, it has paved the way 
for an intensive study of alternative statistical models within the framework of information theory~\cite{Ren}.

Starting from the above premises, in this work
we feature the entangled condensate structure of the vacuum
for mixed fields in the language of nonextensive Tsallis statistics.
We consider Unruh effect as a specific playground. 
In this context, it is shown that the modified Unruh 
distribution for mixed fields can be described by
an appropriately $q-$generalized distribution
based on Tsallis entropy.
This allows us to establish an effective connection
between the nonextensive $q$-entropic index
and the characteristic mixing parameters $\sin\theta$ and $\Delta m$
in a suitable approximation.  We find that $q<1$, 
which corresponds to a superadditive Tsallis entropy.
To make our analysis as transparent as possible, we deal with a
simplified model involving only two scalar
fields. The case of fermion mixing will be shortly 
addressed at the end with similar results.

The remainder of this paper is structured as follows: in Sec.~\ref{UE}
we analyze the canonical quantization of a massive scalar field
for the Rindler observer using the Bogoliubov transformation method. This leads us to derive the Unruh effect in a natural way.
Section~\ref{QFTRIND} contains 
the study of the QFT formalism of flavor mixing
in Rindler spacetime. Exploiting these tools, 
in Sec.~\ref{EDM} we introduce the basics
of nonextensive Tsallis thermostatistics and
investigate the connection with field mixing
at the level of Unruh vacuum condensate.
We discuss our result in relation with the
complex condensate structure acquired by the 
quantum vacuum for mixed fields.
Conclusions and outlook are summarized
in Sec.~\ref{Conc}. The work ends with an Appendix
devoted to review the theory of
field mixing in Minkowski background.

Throughout all the manuscript, we adopt the
mostly negative signature
for the $4$-dimensional metric
and natural units $\hslash=c=k_{\mathrm B}=1$. Furthermore, 
we use the notation
\begin{equation}
\label{notation}
x=\{t,\textbf{x}\},\qquad \bx=\{x^1,\vec{x}\},\qquad \vec{x}=\{x^2,x^3\}\,, 
\end{equation}
for $4$-, $3$- and $2$-vectors, respectively.

\maketitle

\setcounter{equation}{0}

\section{QFT in Rindler spacetime and Unruh effect}
\label{UE}
In this Section we briefly review the quantization of a massive scalar field
for a uniformly accelerated (Rindler) observer. Without loss
of generality, we assume the acceleration to be  
along the $x^1$-axis. In this setting, it is useful to 
introduce the new coordinates $-\infty<\eta,\xi<\infty$, such that
$t=\xi\sinh\eta$, $x^1=\xi\cosh\eta$, while leaving $\vec{x}$ unchanged. 
In terms of these coordinates, Minkowski metric takes the form
\be
\label{Rindme}
ds^2={(dt)}^2-{(dx^1)}^2-\sum_{j=2}^{3}{(dx^j)}^2\underset{\tiny{Rindler\; coord.}}{\longrightarrow}ds^2=\xi^2d\eta^2-d\xi^2-\sum_{j=2}^{3}{(dx^j)}^2\,, 
\ee 
which admits $B=\partial/\partial{\eta}$ as a time-like
Killing vector. Henceforth, we refer to the coordinates 
$\{t,x^1\}$ and $\{\eta, \xi\}$ as Minkowski and Rindler 
coordinates, respectively. Accordingly, the metric~\eqref{Rindme} 
shall be named Rindler metric.

Let us now consider the world-line of fixed
spatial coordinates, i.e.
\be
\label{wl}
\xi(\tau)=a^{-1}=\mathrm{const}\,,\quad \vec{x}(\tau)=\mathrm{const}, 
\ee
where $\tau$ is the proper time measured along the line. 
Substitution of Eq.~(\ref{wl}) into the metric~\eqref{Rindme}
yields $\eta(\tau)=a\tau$, 
i.e., the proper time $\tau$ for an observer moving
along the line (\ref{wl}) 
is the same as the Rindler time $\eta$, 
up to the scale factor $a$. 

\begin{figure}[t]
\resizebox{8.2cm}{!}{\includegraphics{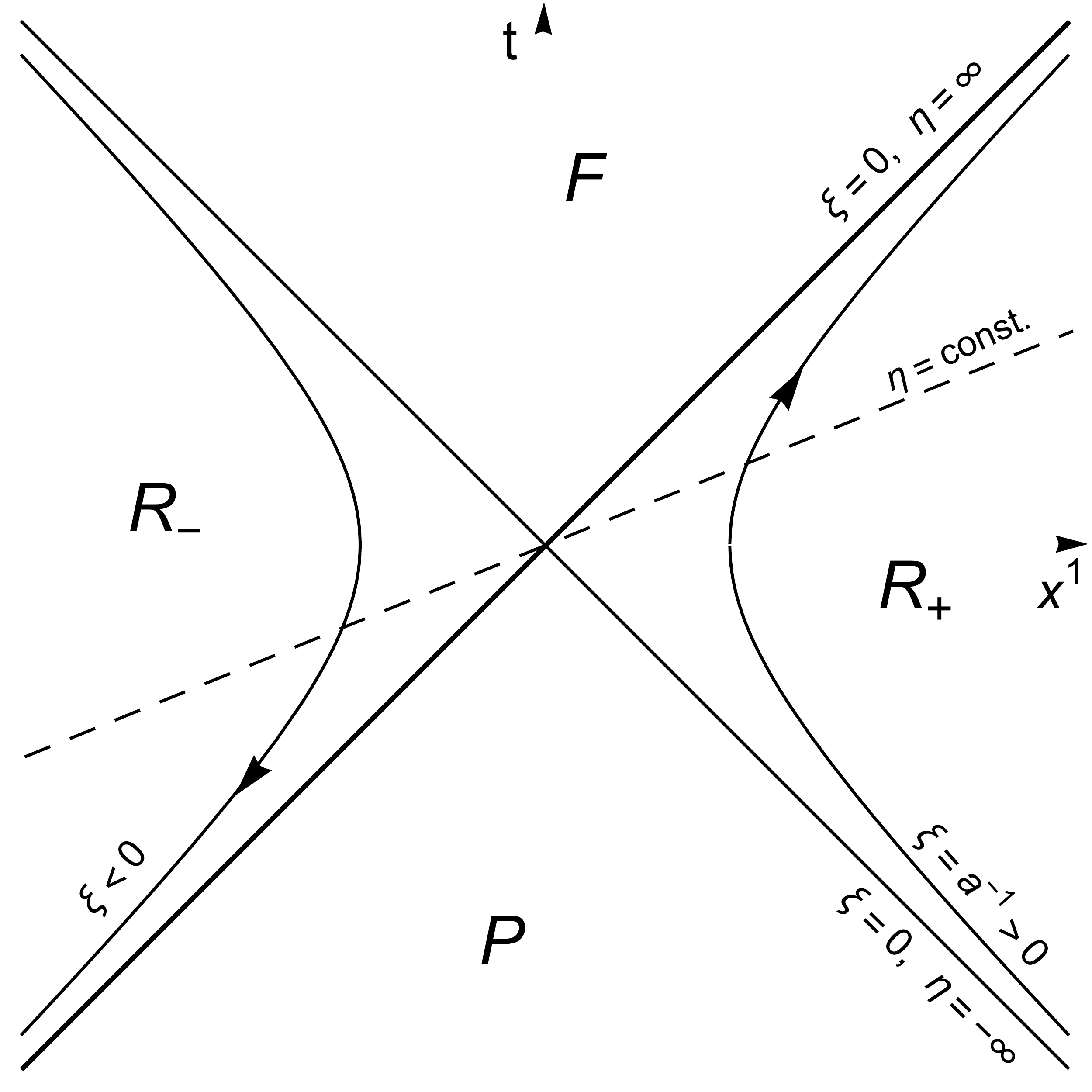}}
\caption{\small{Causal structure of Rindler metric in $1+1$ dimensions. 
We are assuming that Rindler observer accelerates along the $x^1$-axis, 
while the transverse dimensions $x^2, x^3$ are kept constant during the motion. The arrow indicates the direction of motion.}}
\label{figure:Rindler}
\end{figure}

In Minkowski coordinates Eq.~(\ref{wl}) becomes
\begin{equation}
t(\tau)=a^{-1}\sinh a\tau,\quad x^1(\tau)=a^{-1}\cosh a\tau,\quad \vec{x}(\tau)=\mathrm{const}\,, 
\label{minkl}
\end{equation}
which is an hyperbola with asymptotes $t=\pm x^1$
in the $(t,x^1)$ plane. 
In special relativity, it is well known that the hyperbolic motion 
generalizes the concept of Newtonian uniformly accelerated motion, 
with $|a|$ being the magnitude of the proper acceleration~\cite{Mukhanov}.
In particular, for $a>0$ the observer moves along the branch
of hyperbola in the right wedge $R_+=\{x|x^1>|t|\}$, 
while for $a<0$ the motion occurs in the left wedge $R_-=\{x|x^1<-|t|\}$ (see Fig.~\ref{figure:Rindler}). 
This reveals the peculiar features of the causal structure of Rindler spacetime:
since a uniformly accelerated observer in 
$R_+$ cannot receive (send) 
any signal from (to) the future (past) wedge $t > |x^1|$ $(t < -|x^1|)$, 
the asymptote $t=|x^1|$ ($t=-|x^1|$) appears to him
as a future (past) event horizon. Notice that
the time ordering of the two horizons is reversed in $R_-$, 
since the Killing vector $B$ is
past oriented in this wedge.  
Accordingly, a Rindler observer in $R_{+}$ turns out to be
causally separated from $R_{-}$, and vice-versa.  

Bearing in mind the causal structure of the metric~\eqref{Rindme}, 
let us deal with the quantization of a charged scalar 
field $\phi(x)$ of mass $m$ for the Rindler observer\footnote{Where there is no ambiguity, 
we shall denote by $x$ both the sets
of Minkowski and Rindler coordinates.}. 
In Rindler coordinates, the Klein-Gordon equation reads
\begin{equation}
\bigg\{{\bigg(\frac{\partial}{\partial t}\bigg)}^2-\sum_{j=1}^{3}{\bigg(\frac{\partial}{\partial x^j}\bigg)}^2+m^2 \bigg\}\hspace{0.1mm}\,\phi(x)=0\underset{Rindler\;coord.}{\longrightarrow}
\bigg\{\frac{1}{\xi^2}\frac{\partial^2}{\partial\eta^2}-\frac{\partial^2}{\partial\xi^2}-\frac{1}{\xi}\frac{\partial}{\partial\xi}-\sum_{j=2}^{3}{\bigg(\frac{\partial}{\partial x^j}\bigg)}^2+m^2 \bigg\}\hspace{0.1mm}\,\phi(x)=0\,,
\label{KG}
\end{equation}
which has the modes
\be
u_\kappa^{\,(\sigma)}(x)=
\frac{\theta(\sigma\xi)}{2\pi\Omega}\frac{1}{\Gamma(i\Omega)^2\sqrt{\sinh(\pi\Omega)}}\hspace{0.2mm}K_{i\Omega}(\mu_k\xi)\hspace{0.1mm}e^{i\left(\vec{k}\cdot\vec{x}-\sigma\Omega\eta\right)}\,, 
\label{eqn:rindlermodes}
\end{equation}
as solutions of frequency $\Omega>0$ with respect
to the time coordinate $\eta$. Following~\cite{Takagi}, 
here we have introduced the shorthand 
notation $\kappa\equiv(\Omega, \vec{k})$.
The Heaviside step function $\theta(\sigma\xi)$
restricts the support of $u_\kappa^{\,(\sigma)}$ 
to only one of the two Rindler wedges. Specifically, 
$\sigma=+$ refers to the right wedge $R_{+}$, while
$\sigma=-$ to the left wedge. The function 
$\Gamma(i\Omega)$ is the usual Euler's gamma, while $K_{i\Omega}(\mu_k\xi)$ denotes the modified Bessel
function of second kind. The reduced Minkowski frequency is given by  $\mu_k={(m^2+|\vec{k}|^2)}^{1/2}$. 

It is easy to verify that the Rindler modes~\eqref{eqn:rindlermodes} 
form a complete and orthonormal set with respect 
to the Klein-Gordon product in Rindler coordinates. 
This allows us to take
the following expansion for the scalar field~\cite{Takagi}
\begin{equation}
\phi(x)=\sum_{\sigma,\hspace{0.4mm}\Omega}\,\int d^{2}k\,\Big\{{b^{\,(\sigma)}_\kappa}\, u_\kappa^{\,(\sigma)}(x)+{\bar b^{\,(\sigma)\dagger}_\kappa}\, u_\kappa^{\,(\sigma)*}(x) \Big\},
\label{eqn:espanrind}
\end{equation}
where $\sum_{\sigma,\hspace{0.4mm}\Omega}$
stands for $\sum_{\sigma=\pm}\int_{0}^{\infty}d\Omega\hspace{0.2mm}$. The ladder operators ${b^{\,(\sigma)}_\kappa}$ (${\bar b^{\,(\sigma)}_\kappa}$) are assumed to be canonical. They act as annihilators of Rindler particles (antiparticles) 
of frequency $\Omega$ and transverse momentum $\vec{k}$
in the wedge $R_{\sigma}$. Rindler vacuum $|0_{\mathrm{R}}\rangle$ is defined by $b_\kappa^{\,(\sigma)}|0_{\mathrm{R}}\rangle=\bar b_\kappa^{\,(\sigma)}|0_{\mathrm{R}}\rangle=0$, $\forall\sigma, \kappa\,$. On the other hand, ${b^{\,(\sigma)\dagger}_\kappa}$ (${\bar b^{\,(\sigma)\dagger}_\kappa}$) creates a Rindler particle (antiparticle) with the same quantum numbers as defined above.

We now focus on the relation between
the quantization~\eqref{eqn:espanrind} and 
the Minkowski plane-wave expansion
\be
\phi(x)=\int d^{3}{k}\, \Big\{a_{\textbf{k}}\, \uuu_{\textbf{k}}(x)+ {\bar a_\textbf{k}}^\dagger\, \uuu_\textbf{k}^{\hspace{0.1mm}*}(x) \Big\},
\label{eqn:expans0}
\end{equation}
where $U_\bk(x)={[{2\omega_{\bk}{(2\pi)}^{3}}]}^{-1/2}e^{i\left(\bk\cdot\bx-\omega_{\bk} t\right)}$ are the plane-waves of frequency
$\omega_{\textbf{k}}=({m^2+|\textbf{k}|^2})^{1/2}$. Here, 
$a_{\bk}$ ($\bar a_{\bk}$) denote
the canonical annihilators of 
Minkowski particles (antiparticles) with momentum
$\bk$ and frequency $\omega_{\bk}$, such that
$a_\bk|0_{\mathrm{M}}\rangle=\bar a_\bk|0_{\mathrm{M}}\rangle=0$, $\forall \bk$,
where $|0_{\mathrm{M}}\rangle$ is the Minkowski vacuum.
As before, the r\^ole of creation operators
is played by the adjoint $a^\dagger_\bk$ ($\bar a^\dagger_\bk$). 
Since this formalism holds
for the set of inertial observers in Minkowski spacetime, 
in what follows we indifferently refer to the expansion~\eqref{eqn:expans0} as either Minkowski or inertial field
quantization.

To derive the Bogoliubov transformation 
between the two field representations introduced above, we
compare the expansions~\eqref{eqn:espanrind} 
and~\eqref{eqn:expans0}
on a space-like hypersurface $\Sigma$ 
which lies in the Rindler manifold $R_\pm$.
By multiplying both sides by the Rindler mode $u^{(\sigma)}_{\kappa}$,
we obtain~\cite{Takagi}
\begin{equation}
b_{\kappa}^{\,(\sigma)}=\int d^3k'\left\{a_{\textbf{k}'}\hspace{0.1mm}(u^{(\sigma)}_{\kappa},U_{\textbf{k}'})\,+\,\bar a^\dagger_{\textbf{k}'}\hspace{0.1mm}(u^{(\sigma)}_{\kappa},U^*_{\textbf{k}'})
\right\}\,. 
\label{bogo}
\end{equation}
The explicit expressions of the Bogoliubov coefficients $(u^{(\sigma)}_{\kappa},U_{\textbf{k}'})$ and $(u^{(\sigma)}_{\kappa},U^*_{\textbf{k}'})$
are rather awkward to exhibit. They are given in~\cite{Takagi}.
However, Eq.~\eqref{bogo}
can be cast in a more transparent form 
by introducing the following superposition of $a_{\textbf{k}}$-operators
\begin{equation}
\label{eqn:operat-d}
d_{\kappa}^{\,(\sigma)}=\int dk_1\, \frac{1}{\sqrt{2\pi\omega_\bk}}\, {\bigg(\frac{\omega_{\bk}+k_1}{\omega_{\bk}-k_1}\bigg)}^{i\sigma\Omega/2}\, a_{\bk},
\end{equation}
(similarly, the definition of $\bar d_{\kappa}^{\,(\sigma)}$ is obtained
by replacing $a_{\bk}$ with $\bar a_{\bk}$). In~\cite{Takagi},
it has been shown that these new operators still obey the canonical commutator. Furthermore, they share a common vacuum with the 
$a_{\textbf{k}}$'s, since they are linear combinations of these annihilators only. In other terms, the $d$-operators
provide an alternative representation
for the scalar field, which is unitarily equivalent
to the plane-wave quantization from the perspective
of inertial observers\footnote{At first glance, the physical meaning
of $d$-operators may appear quite unclear.
However, in~\cite{Takagi} it has been shown that 
they diagonalize the generator of Lorentz boosts along the $x^1$-axis. Therefore, the field quantization in this representation 
exploits the symmetry of Minkowski spacetime under boost transformations, just as the plane-wave and spherical-wave quantizations 
rely on the symmetry under spacetime translations and rotations, respectively.}. This has been discussed in more detail in~\cite{Luciano}.

In terms of the operators~\eqref{eqn:operat-d}, 
Eq.~\eqref{bogo} can be rewritten as
\be
\label{nbt}
b_{\kappa}^{\,(\sigma)}=\sqrt{1+N_{\mathrm{BE}}(\Omega)}\hspace{0.1mm}d_{\kappa}^{\,(\sigma)}\,+\,\sqrt{N_{\mathrm{BE}}(\Omega)}\hspace{0.1mm}\bar d_{\tilde\kappa}^{\,(-\sigma)\,\dagger}\,,
\ee
where $\tilde\kappa\equiv(\Omega,-\vec{k})$ and
\be
\label{therm}
N_{\mathrm{BE}}(\Omega)=
\frac{1}{e^{2\pi\Omega}-1}\,.
\ee
is the Bose-Einstein distribution function.

Next, by resorting to Eqs.~\eqref{eqn:operat-d} and~\eqref{nbt}, we can evaluate the expected number spectrum of Rindler particles in the
Minkowski vacuum, obtaining 
\be
\label{parcon}
\langle0_{\mathrm{M}}|b_{\kappa'}^{\,(\sigma')\dagger}\hspace{0.2mm}b_{\kappa}^{\,(\sigma)}|0_{\mathrm{M}}\rangle=N_{\mathrm{BE}}(\Omega)\hspace{0.2mm}\delta_{\sigma\sigma'}\hspace{0.2mm}\delta^3(\kappa-\kappa')\,,
\ee
(similarly for $\bar b_{\kappa}^{\,(\sigma)}$).
Since the proper energy of quanta detected by
an observer moving with uniform acceleration $a$
is 
\be 
\label{EnRind}
E=a\Omega\,,
\ee
it is more appropriate to rewrite the distribution~\eqref{therm}
in the form
\be
\label{newform}
N_{\mathrm{BE}}(\Omega)=\frac{1}{e^{E/T_{\mathrm{U}}}-1}\,,
\ee
where $T_{\mathrm{U}}=a/(2\pi)$ is the 
Unruh temperature~\cite{Unruh}. 
Thus, we recover the well-known
Unruh result that Minkowski vacuum appears
as a Bose-Einstein thermal distribution of Rindler particles, 
with temperature $T_{\mathrm{U}}$ being 
proportional to the magnitude of the proper
acceleration~\cite{Unruh}. 

As discussed in~\cite{Takagi}, 
the inherent properties of the 
condensate~\eqref{parcon} can be 
explored in depth 
by expressing the Minkowski vacuum $|0_\mathrm{M}\rangle$
in terms of the Rindler one $|0_\mathrm{R}\rangle$. 
In so doing, we infer that the inertial vacuum acquires the structure of 
a coherent state of pairwise-correlated Rindler particles. Specifically, 
an excitation in the positive wedge $R_+$ is 
correlated to an excitation of opposite spatial momentum 
in the negative region $R_+$, and vice-versa. 
Since the two wedges are causally disconnected, 
this turns out to be an EPR-like correlation
between space-like separated quanta.

The spectrum~\eqref{parcon} diverges
for any fixed $\kappa=\kappa'$. 
This is due to the fact that the creation operators
$a_{\bk}^\dagger$, $b_{\kappa}^\dagger$
(and the corresponding operators for
antiparticles) do not produce normalizable
states when applied on the respective vacua.
To avoid conceptual difficulties we shall otherwise 
encounter later, we introduce the following set of functions~\cite{Hawk1995}
\be
\label{fj}
f_{nl}(k)=\chi_n(k)\exp^{-2\pi i l k/\varepsilon}\,,
\ee
\be
\chi_n(k)=\left\{\begin{array}{rcl}
&&\hspace{-4mm}\varepsilon^{-1/2}\,\,\,\,\,\,\,\,\, \mathrm{for} \left(n-\frac{1}{2}\right)\varepsilon<k<\left(n+\frac{1}{2}\right)\varepsilon\\[3mm]
&&\hspace{-4mm}0 \,\,\,\,\,\,\,\,\,\,\,\,\,\,\,\,\,\,\,\,\mathrm{otherwise}
\end{array}\right.\,,
\ee
where $\varepsilon$ is a positive constant of dimension
of inverse length. By exploiting the completeness and orthonormality 
of this set~\cite{Takagi}, we define
the Minkowski wave packet by
\be
\label{Mwp}
U_{n l \vec{k}}(x)=\int dk_1 f_{nl}(k_1)U_{\bk}(x)\,,
\ee
where $n$ and $l$ run over all the integers. 

On the other hand, to form the Rindler wave packet,
we restrict the subscript $n$ of $\{f_{nl}\}$ to 
positive integers. Notice that this 
does not affect the orthonormality nor
the completeness of the set~\eqref{fj}, 
provided that the argument $k$ is now replaced
by the Rindler frequency $\Omega$ and
$\varepsilon$ is assumed to be dimensionless. 
We then define
\be
\label{Rwp}
u_{n l \vec{k}}(x)=\int_{0}^{\infty}d\Omega f_{nl}(\Omega)u^{(\sigma)}_{\kappa}(x)\,. 
\ee
Two comments are in order here: first, the wave packets~\eqref{Mwp}
and~\eqref{Rwp} satisfy a box-like normalization
and are complete, due to the definition~\eqref{fj} of the smearing
function. Furthermore, in the above construction we
have left the reduced momentum $\vec{k}$ untouched, 
as it does not enter the Bose-Einstein distribution function~\eqref{newform}
explicitly. More properly, we should extend
the wave packet formalism to this quantum number as well,  
resulting in a new pair of subscripts in place of each 
component of $\vec{k}$. However, 
since this procedure would burden the
notation without providing any conceptual advantage, 
we continue to use the symbol $\vec{k}$, 
taking care of this aspect.

We can now repeat the computation of the
number spectrum of Rindler particles in the Minkowski
vacuum. By choosing the parameter $\varepsilon$
in the definition of $f_{nl}(k_1)$ much less than the
reduced Minkowski frequency $\mu_k$ and 
the $\varepsilon$ in $f_{nl}(\Omega)$ much less than unity, 
we are led to~\cite{Takagi}
\be
\label{wpU}
\langle0_{\mathrm{M}}|\hspace{0.2mm}b_{n'l'\vec{k}'}^{\,(\sigma')\dagger}\hspace{0.1mm}b_{nl\vec{k}}^{\,(\sigma)}\hspace{0.2mm}|0_{\mathrm{M}}\rangle=N_{\mathrm{BE}}({\Omega_n})\delta_{\sigma\sigma'}\delta_{nn'}\delta_{ll'}\delta_{\vec{k}\vec{k'}}\,,
\ee
where $\Omega_n=n\varepsilon$. Of course, 
for fixed values of $\sigma=\sigma', n=n', l=l', m=m'$ and $\vec{k}=\vec{k}'$, this gives
\be
\label{wpUbis}
\mathcal{N}(\Omega_n)\equiv\langle0_{\mathrm{M}}|\hspace{0.2mm}b_{nl\vec{k}}^{\,(\sigma)\dagger}\hspace{0.1mm}b_{nl\vec{k}}^{\,(\sigma)}\hspace{0.2mm}|0_{\mathrm{M}}\rangle=N_{\mathrm{BE}}({\Omega_n})\,,
\ee 
As expected, the use
of properly normalizable wave packets results into a 
regularization of the Unruh spectrum~\eqref{parcon}. In the next Section, 
we shall see how the distribution~\eqref{wpUbis} 
is modified when dealing with the superposition of fields
with different masses.

\section{QFT of flavor mixing in Rindler spacetime}
\label{QFTRIND}
The QFT treatment of flavor mixing, originally developed
for Dirac neutrinos in Minkowski background~\cite{BV95} 
and later extended to mesons, such as the
$K^0-\bar K^{0}$, $B^0-\bar B^0$ and $\eta-\eta'$ systems~\cite{BlasCap,Ji}, has revealed a series of nontrivial features that are totally missed by quantum mechanics. 
These aspects are reviewed in the Appendix,  
with particular emphasis on the issue of the 
unitary inequivalence between the Fock space for fields with definite flavor and the Fock space for fields
with definite mass.  Following~\cite{Luciano}, here we generalize the quantization 
of mixed fields to the Rindler metric. 
We consider the mixing transformations in a simplified two-flavor 
model with charged scalar fields\footnote{Strictly speaking, 
for bosons we should refer to the mixing of quantum numbers such as the strangeness or isospin, rather than flavor.
However, in what follows we improperly label
such a quantum number as flavor and the corresponding 
fields as definite flavor fields. On the other hand, the fields with definite
mass will be referred to as mass fields.}. Denoting by $\chi=A,B$ ($i=1,2$)
the flavor (mass) label, these transformations read
\bea 
\label{Ponteca}
\phi_{A}(x)&=& \phi_{1}(x)\, \cos\theta  + \phi_{2}(x)\, \sin\theta\,,\\ [2mm]
\label{Pontecb}
\phi_{B}(x) &=&-  \phi_{1}(x)\, \sin\theta   +  \phi_{2}(x)\, \cos\theta\,,
\eea
where $\theta$ is the mixing angle and
$\phi_i$ ($i=1,2$) are two free charged scalar fields
of masses $m_i$, such that $m_2\neq m_1$.  For definiteness, 
we set $m_2>m_1$. Let $\pi_i=\partial_t\phi_i^\dagger$
be the conjugate momenta.

In the canonical quantization formalism, it is known that
\begin{equation}
\qquad\big[\phi_i(x),\,\pi_j(x')\big]_{t=t'}=\big[\phi^\dagger_i(x),\,\pi^\dagger_j(x')\big]_{t=t'}=i\,\delta^3(x-x')\,\delta_{ij}\,,\qquad i,j=1,2\,,
\label{eqn:relcommcan}
\end{equation}
with all other equal-time commutators vanishing.
In the Appendix we discuss the algebraic structure
of Eqs.~\eqref{Ponteca} and~\eqref{Pontecb}, showing that each of them
appears as a rotation combined with a Bogoliubov transformation 
when seen at level of ladder operators.
Notice that this peculiar structure, which is absent in quantum mechanics, arises from the necessity to take account of the
antiparticle degrees of freedom intrinsically 
built in QFT. 
As a result, the vacuum state for flavor fields
becomes a condensate of massive particle-antiparticle pairs (see Eq.~\eqref{eqn:denscondbosoni2}). 

To find out how the phenomenon of mixing
appears to the Rindler observer, we retrace
the same steps leading to Eq.~\eqref{eqn:espanrind}
and consider the following free
field-like expansions for mixed fields~\cite{Luciano}
\be
\label{fRind}
\phi_\ell(x)=\sum_{\sigma,\hspace{0.4mm}\Omega}\,\int d^{2}k\,\Big\{{b^{\,(\sigma)}_{\kappa,\ell}}(\eta)\, u_{\kappa,j}^{\,(\sigma)}(x)+{\bar b^{\,(\sigma)\dagger}_{\kappa,\ell}}(\eta)\, u_{\kappa,j}^{\,(\sigma)*}(x) \Big\},\quad (\ell,j)=\left\{(A,1),(B,2)\right\},
\ee
where we have used the shorthand notation 
$b^{\,(\sigma)}_{\kappa,\ell}(\eta)\equiv b^{\,(\sigma)}_{\kappa,\ell}(\theta, \eta)$ for the ladder operators (similarly for $\bar b^{\,(\sigma)}_{\kappa,\ell}(\eta)$). By comparison with Eq.~\eqref{eqn:dynamicalmap2}, it is clear that these operators 
provide the Rindler counterpart of
Minkowski flavor annihilators given 
in Eq.~\eqref{tra}. It is a matter of calculations
to show that they obey the (equal-time) canonical commutators. 

As remarked above, 
the mixing relations at level of ladder operators
hide a Bogoliubov transformation 
between the flavor and mass bases. 
At the same time, the field quantizations 
for Minkowski and Rindler observers are connected
to each other by the Bogoliubov transformation~\eqref{nbt}
responsible for the thermal Unruh effect. Overall, 
we expect that the Rindler annihilators in the flavor
representation are related to the corresponding Minkowski 
operators in the mass basis by a combination of these 
two Bogoliubov transformations.  To analyze such an interplay, 
we compare the expansions~\eqref{fRind} and~\eqref{eqn:dynamicalmap2}. By using Eqs.~\eqref{Ponteca}-\eqref{Pontecb} and the transformation~\eqref{tra}, 
after some tedious but straightforward calculations, we obtain~\cite{Luciano}
\begin{eqnarray}
\label{eqn:bogtransfbis}
b_{\kappa,A}^{\,(\sigma)}&=&\sqrt{1+N_{\mathrm{BE}}(\Omega)}\,\,
\Big[\cos\theta\, d_{\kappa,1}^{\,(\sigma)}
\,+\,\sin\theta\hspace{-0.6mm}\sum_{\hspace{0.6mm}\sigma',\hspace{0.1mm}\Omega'}\Big(d_{(\Omega',\vec{k}),2}^{\,(\sigma')}\,\;\bogocoeffAlpha \,+\,\bar d_{(\Omega',-\vec{k}),2}^{\,(\sigma')\dagger}\,\;\bogocoeffBeta\,\Big)\Big]\\[2mm]
\nonumber
&+&\,\sqrt{N_{\mathrm{BE}}(\Omega)}\,\,\Big[\cos\theta\,\bar d_{\tilde\kappa,1}^{\,(-\sigma)\dagger}
\,+\,\sin\theta\hspace{-0.6mm}\sum_{\hspace{0.6mm}\sigma',\hspace{0.1mm}\Omega'}\Big(\bar d_{(\Omega',-\vec{k}),2}^{(-\sigma')\dagger}\,\;\bogocoeffAlphamensigstarkapdue\,+\,d_{(\Omega',\vec{k}),2}^{(-\sigma')}\,\;\bogocoeffBetamensigstarkapdue\,\Big)\Big],
\end{eqnarray}
where we have omitted for simplicity the time-dependence
of $b_{\kappa,A}^{\,(\sigma)}$ (a similar expression hold true 
for $b_{\kappa,B}^{\,(\sigma)}$ as well). 
We stress that the $d$-operators
are an equivalent way of rewriting the standard 
Minkowski annihilators appearing in Eq.~\eqref{eqn:expans0} 
(see Eq.~\eqref{eqn:operat-d}).
From the above relation, we can clearly   
distinguish the action of the thermal
Bogoliubov transformation (encoded by the coefficients
${(1+N_{\mathrm{BE}})^{1/2}}$ and ${N^{1/2}_{\mathrm{BE}}}$) from the action of the mixing Bogoliubov transformation
(which appears through the coefficients $\bogocoeffAlpha$
and $\bogocoeffBeta$). The Bogoliubov coefficients related to mixing are given by~\cite{Luciano}
\begin{eqnarray}
&& \bogocoeffAlpha =\int^{+\infty}_{-\infty}\frac{dk_1}{4\pi}\lf(\frac{1}{\omega_{\bk,1}}+\frac{1}{\omega_{\bk,2}}\ri){\lf(\frac{\omega_{\bk,1}+k_1}{\omega_{\bk,1}-k_1}\ri)}^{i\sigma\Omega/2} {\lf(\frac{\omega_{\bk,2}+k_1}{\omega_{\bk,2}-k_1}\ri)}^{-i\sigma'\Omega'/2}
e^{i(\omega_{\bk,1}-\,\omega_{\bk,2})t}\,,
\label{eqn:coefficientunobis}
\\ [4mm]
&& \bogocoeffBeta =\int^{+\infty}_{-\infty}\frac{dk_1}{4\pi}\lf(\frac{1}{\omega_{\bk,2}}
-\frac{1}{\omega_{\bk,1}}\ri){\lf(\frac{\omega_{\bk,1}+
k_1}{\omega_{\bk,1}-k_1}\ri)}^{i\sigma\Omega/2} {\lf(\frac{\omega_{\bk,2}+k_1}{\omega_{\bk,2}-k_1}\ri)}^{-i\sigma'\Omega'/2}
e^{i(\omega_{\bk,1}+\,\omega_{\bk,2})t}\,.
\label{eqn:coefficientduebis}
\end{eqnarray}
Despite the nontrivial structure of $\bogocoeffAlpha$
and $\bogocoeffBeta$, interesting implications of Eq.~\eqref{eqn:bogtransfbis} can still be derived for $t=\eta=0$
and in the reasonable approximation of small difference 
between the masses of the two fields, i.e. $|\Delta m^2|/{m_1^2}=|m_2^2-m_1^2|/m_1^2\ll 1$.\footnote{Notice that the assumption $|\Delta m^2|/m^2\ll 1$ makes even more physical sense if considered for mixing of neutrino fields, the study of which is reserved for future investigation.}
Indeed, if we evaluate the number spectrum of mixed particles detected by the Rindler observer in the inertial vacuum, we get to the leading order~\cite{Luciano}  
\be
\langle0_\mathrm{M}|\,b_{\kappa,\chi}^{\,(\sigma)\dagger}(0)\,
b_{\kappa',\chi}^{\,(\sigma)}(0)\,|0_\mathrm{M}\rangle\,
=\, N_{\mathrm{BE}}(\Omega)\,\delta^3(\kappa-\kappa')-\,\frac{|\Delta m^2|}{8\,\mu_{k,1}^{\hspace{0.1mm}2}}\sin^2\theta\,\frac{\sigma\,(\Omega'+\Omega)\,G(\Omega,\Omega')}{\sinh\left[\frac{\pi}{2}\,\sigma\,(\Omega'+\Omega)\right]}\,\delta(\vec{k}-\vec{k}')\,+\,\mathcal{O}\left(\frac{|\Delta m^2|}{\mu^2_{k,1}}\right)^2\,,
\label{nontherm}
\ee
for $\chi=A,B$, where
\be
G(\Omega,\Omega')\equiv\sqrt{1+N_{\mathrm{BE}}(\Omega)}\,\sqrt{N_\mathrm{BE}(\Omega')}\,+\,\sqrt{N_{\mathrm{BE}}(\Omega)}\,\sqrt{1+N_{\mathrm{BE}}(\Omega')}\,.
\label{eqn:G}
\ee
Thus, the standard Bose-Einstein distribution
of Unruh vacuum condensate~\eqref{parcon} 
turns out to be spoilt in the presence of 
mixed fields. Notice that, for $\theta\rightarrow 0$,  Eq.~(\ref{nontherm}) recovers the usual result, as expected in the absence of mixing. 
Similar considerations hold true for $m_1\rightarrow m_2$ and in the relativistic limit $|\vec{k}|^{2}\gg m_{1}^2+m_{2}^2$, since the parameter ${|\Delta m^2|}/{\mu^2_{k,i}}$ becomes increasingly small.

As argued at the end of Sec.~\ref{UE}, 
in order to avoid unphysical divergencies in the spectrum, 
it is convenient to revisit the above formalism
by employing wave packets. With the aid
of Eqs.~\eqref{fj}-\eqref{Rwp}, the 
density~\eqref{wpUbis} of Rindler mixed particles 
with quantum numbers $\Omega_n, l, \vec{k}$
then becomes 
\be
\label{ntbis}
\mathcal{N}_{\theta,\Delta m}(\Omega_n)\,\equiv\,\langle0_{\mathrm{M}}|\hspace{0.2mm}b_{nl\vec{k}}^{\dagger}(0)\hspace{0.1mm}b_{nl\vec{k}}(0)|0_{\mathrm{M}}\rangle
\,=\, N_{\mathrm{BE}}({\Omega_n})\,-\,\frac{|\Delta m^2|}{4\,\mu_{k,1}^{\hspace{0.1mm}2}}\sin^2\theta\,\frac{\Omega_{n}\,G(\Omega_n,\Omega_n)}{\sinh\left({\pi}\Omega_n\right)}\,+\,\mathcal{O}\left(\frac{|\Delta m^2|}{\mu^2_{k,1}}\right)^2\,,
\ee
where we have dropped the indices $\sigma$
and $\chi$ from $\langle0_{\mathrm{M}}|\hspace{0.2mm}b_{nl\vec{k}}^{\dagger}(0)\hspace{0.1mm}b_{nl\vec{k}}(0)|0_{\mathrm{M}}\rangle
$, since the r.h.s. of  Eq~\eqref{nontherm}
is in fact insensitive to them. 
By resorting to the definition~\eqref{eqn:G} of $G(\Omega,\Omega)$, 
we are finally led to
\be
\label{newmodBE}
\mathcal{N}_{\theta,\Delta m}(\Omega_n)\,=\, N_{\mathrm{BE}}({\Omega_n})\,-\,\frac{|\Delta m^2|}{4\,\mu_{k,1}^{\hspace{0.1mm}2}}\sin^2\theta\,\Omega_n \csch^2(\pi\Omega_n)\,+\,\mathcal{O}\left(\frac{|\Delta m^2|}{\mu^2_{k,1}}\right)^2\,,
\ee
where $\csch(x)=1/\sinh(x)$. 

In~\cite{Luciano} the modified distribution~\eqref{nontherm} (or, equivalently,~\eqref{newmodBE}) has been interpreted as being due
to a breakdown of the thermality of Unruh radiation 
induced by field mixing. We further elaborate
on the physical meaning of the result~\eqref{newmodBE} in the next Section.

\section{Flavor Mixing and q-generalized Tsallis statistics}
\label{EDM}
In the standard Boltzmann-Gibbs thermodynamics, 
it is well known that entropy is 
an additive quantity, which means that, given
two probabilistically independent systems $A$ and $B$ with 
entropies $S_{\mathrm{BG}}(A)$ and $S_{\mathrm{BG}}(B)$, respectively, the total entropy is simply $S_{\mathrm{BG}}(A+B)=S_{\mathrm{BG}}(A)+S_{\mathrm{BG}}(B)$. 
At the statistical level, the Boltzmann-Gibbs 
entropy of a system in an equilibrium macrostate
can be expressed in terms of the corresponding
microscopic configurations as
\be
\label{BGent}
S_{\mathrm{BG}}=-\sum_{i=1}^Wp_i\log p_i\,,
\ee
for a set of $W$ discrete microstates, where
$\{p_i\}$ is the set of probability distribution with 
the condition $\sum_{i=1}^{W}p_i=1$. 
If probabilities are all equal, this takes the well-known form $S_{\mathrm{BG}}=\log W$. It is immediate
to check that $S_{\mathrm{BG}}$ satisfies
the additivity property as defined above. 

Despite the wide range of applicability of the Boltzmann-Gibbs theory,
for complex systems exhibiting long-range
interactions and/or spacetime entanglement,  
it has been argued that the standard 
Boltzmann-Gibbs theory should be generalized
to a nonextensive statistical mechanics
based on the nonadditive Tsallis entropy~\cite{Tsallis1,Tsallis2,Tsallis3,Tsallis4}
\be
\label{TE}
S_q\,=\,\frac{1-\sum_{i=1}^Wp_i^q}{q-1}\,=\,\sum_{i=1}^Wp_i\log_q \frac{1}{p_i}\,,
\ee
with
\be
\log_q z \equiv \frac{z^{1-q}-1}{1-q}, \quad (\log_1 z=\log z)\,.
\ee
Note that $S_q$ recovers Boltzmann-Gibbs entropy 
$S_{\mathrm{BG}}$ in the $q\rightarrow1$ limit.
Furthermore, by considering again two probabilistically
independent systems such that $p_{ij}^{A+B}=p_{i}^Ap_{j}^B, 
\, \forall {(i,j)}$, the definition~\eqref{TE} leads to
\be
\label{sqab}
S_{q}(A+B)=S_q (A) + S_q (B) + (1-q) S_q (A) S_q(B)\,, 
\ee
indicating that $S_{q}$ is superadditive or subadditive, depending
on whether $q<1$ or $q>1$. Thus, the dimensionless
index $q\in\mathbb{R^+}$ quantifies the departure
of Tsallis entropy from 
Boltzmann-Gibbs one. For this reason, it is named
nonextensive Tsallis parameter. Paradigmatic 
examples of systems obeying the generalized
statistics~\eqref{TE} are the strongly
gravitating black holes~\cite{Tsallis3}, 
albeit in recent years Tsallis thermostatistics 
has found applications in a variety of physical scenarios~\cite{App1,App3,App7,App8,App11,App13,App14}.

Now, within an approximation called factorization approach, 
it has been shown that the Tsallis entropy~\eqref{TE}
can be used to derive the following generalized Bose-Einstein 
distribution~\cite{Buyu,Buyu2,Buyu3,Buyu4,Chen}
\be
\label{modBE}
N_q(\epsilon_n)=\frac{1}{\left[1+(q-1)\beta \epsilon_n\right]^{1/(q-1)}-1}\,,
\ee
where $\epsilon_n$ is the energy of the
$n$-th state of the system and $\beta=1/T$. 
Clearly, for $q\rightarrow1$, Eq.~\eqref{modBE}
gives back the conventional Bose-Einstein distribution. 
By definition, the generalized distribution $N_q$
must be non-negative. This gives rise to the
following constraints
\be
\label{constraint}
\left\{\begin{array}{rcl}
&&\hspace{-4mm}0\le\epsilon_n\le \left[(1-q)\beta\right]^{-1}\,\,\,\,\,\,\,\,\,\ \mathrm{for}\,\,\, q<1,\\[3mm]
&&\hspace{-4mm}\epsilon_n\ge0\hspace{3cm}\mathrm{for}\,\,\, q>1\,.
\end{array}
\right.
\ee
For the sake of clarity, it must be said that
Eq.~\eqref{modBE} can only be regarded as an approximation~\cite{AppBE}. Indeed, 
the exact generalized distribution cannot be derived analytically for arbitrary values of $q$. However, for systems with a relatively large total number of particles (such as fields), the difference between the exact and approximated expressions
turns out to be fairly negligible at very low temperatures (see~\cite{AppBE} for more detailed numerical estimations). Hence, since typical values
of Unruh temperatures are expected to be extremely small
(we recall that an acceleration $a\simeq10^{20}\,\mathrm{m/s^2}$ 
is barely enough to reach a temperature of $1\,{\mathrm{K}}$), 
we safely fall within the regime of validity of Eq.~\eqref{modBE}, which can then be considered as the starting point of our next computations. 

In the previous Section we have emphasized
that Unruh spectrum for mixed fields loses its 
characteristic Planckian profile, the deviation
being proportional to the mixing parameters
(see Eq.~\eqref{newmodBE}). Given the complex
entangled structure induced by mixing in the vacuum state, 
the question naturally arises as to whether such an effect 
can be explained in mechanical statistical terms
by resorting to the nonextensive Tsallis framework.
Of course, since the correction in Eq.~\eqref{newmodBE} slightly affects the Bose-Einstein spectrum at both high and low energy regimes, 
it is reasonable to expand the generalized 
distribution~\eqref{modBE} for tiny departures 
of $q$ from unity. To the leading order, we obtain
\begin{eqnarray}
\label{Loq}
N_q(\epsilon_n)=\frac{1}{e^{\beta \epsilon_n}-1}\,+\,\frac{1}{8}\left(\beta \epsilon_n\right)^2\,\csch^2\left(\frac{\beta \epsilon_n}{2}\right)\left(q-1\right)\,+\,\mathcal{O}\left(q-1\right)^2.
\end{eqnarray}
To compare with the distribution function~\eqref{newmodBE}, 
we resort to Eq.~\eqref{EnRind} and set $\epsilon_n=a\Omega_n$, 
$\beta=1/T_{\mathrm{U}}=2\pi/a$, where
the Unruh temperature $T_{\mathrm{U}}$ has been
defined after Eq.~\eqref{newform}. By plugging into 
$N_q(\epsilon_n)$, this becomes
\be
\label{Loq2}
N_{q}(\Omega_n)=N_{\mathrm{BE}}(\Omega_n)\,+\,\frac{\pi^2}{2}\,\Omega_n^2\,\csch^2\left(\pi\hspace{0.3mm}\Omega_n\right)\left(q-1\right)\,+\,\mathcal{O}\left(q-1\right)^2\,,
\ee
where the zeroth-order term $N_{\mathrm{BE}}$
is the distribution function~\eqref{newform}.
Therefore, within Tsallis thermostatistics, the 
distribution which extremizes the entropy~\eqref{TE}
according to the maximum entropy principle 
can be expanded around $q=1$ as above. 
At a conceptual level, we notice that 
significant deviations from the Bose-Einstein
spectrum still arise at the lowest order, since
the extra term depends on the energy
scale in a nontrivial way. 

To show the correspondence between
the modification induced by flavor mixing
and the $q$-generalized distribution based on 
Tsallis entropy, let us now compare 
Eqs.~\eqref{newmodBE} and~\eqref{Loq2}. 
A straightforward calculation gives (to the leading order)
\be
\mathcal{N}_{\theta,\Delta m}(\Omega_n)=N_{q}(\Omega_n) \,\,\,\Longrightarrow\,\,\, q=1-\frac{|\Delta m^2|}{2\pi^2\mu_{k,1}^{\hspace{0.1mm}2}\hspace{0.1mm}\Omega_n }\sin^2\theta\,, 
\label{q1}
\ee
which implies $q<1$, $\forall \, \Omega_n$. This means that we are in the superadditive regime of 
Tsallis entropy, as it can be seen from Eq.~\eqref{sqab}. 

Thus, the thermostatistical properties of 
Unruh condensate for mixed particles can
be effectively described in terms of the nonextensive Tsallis statistics, 
the entropic $q$-index satisfying the condition~\eqref{q1}. 
As expected, the deviation of $q$
from unity depends on the mixing angle
and the mass difference in such a way
that, for $\theta\rightarrow0$ and/or $\Delta m\rightarrow0$, 
the usual Boltzmann-Gibbs theory with 
$q=1$ is recovered. 

The above considerations provide us with
an alternative way of interpreting the modified distribution~\eqref{newmodBE}.
Indeed, in~\cite{Luciano} mixing 
was seen as the origin of a breakdown of
the thermality of Unruh effect via the appearance
of exotic terms in the spectrum. 
On the other hand, the present result shows that
one can still maintain the standard thermal picture, 
provided
that the underlying statistics is assumed to obey Tsallis's prescription. 
In passing, we point out
that a similar analysis has been developed in~\cite{Shababi}
in the context of deformations of Heisenberg uncertainty
principle (Generalized Uncertainty Principle).  
Even in that case it has been argued that 
GUP corrections to Unruh effect for a gas of relativistic
massive particles can be
mimicked by a Tsallis-like statistics with 
a modified ($q$-dependent) formula
for Unruh temperature. Connections between
deformed uncertainty relations and generalized
entropies in the framework of Unruh effect have also been discussed in~\cite{deformed2}.

A remarkable property to comment on 
is the running behavior of $q$ as a function
of the energy scale $\Omega_n$. 
Although not envisioned by Tsallis in his original approach, 
this should not be entirely surprising for quantum field theoretical or quantum gravity systems when renormalization group is applied~\cite{App13}. 
A similar scenario with a varying nonextensive parameter 
has been recently discussed in~\cite{App13}
in the context of modified cosmological models.

In this regard, one might spot a 
pathological behavior of Eq.~\eqref{q1} 
in the limit of vanishing Rindler frequency $\Omega_n$. 
Actually, it must be stressed that 
modes with frequency below 
a certain threshold lie outside the domain of validity
of the approximation~\eqref{Loq}, and thus of our analysis. 
Indeed, for $\Omega_n\rightarrow0$
the $q$ index strongly deviates from unity, 
which \emph{a posteriori} would invalidate 
the series expansion~\eqref{Loq2}. In order
to keep our formalism self-consistent, the condition 
$|q-1|\ll1$ must be satisfied, which in turn implies
the cutoff $\Omega_n\gg|\Delta m^2|\sin^2\theta/(2\pi^2\mu^{\hspace{0.1mm}2}_{k,1})$. 
This means that the more accurate the approximation
of small mass difference between the mixed fields, 
the higher the number of $\Omega_n$-frequency modes 
that fit with the $q$-generalized Bose-Einstein
distribution~\eqref{modBE}. In the 
limit $|\Delta m^2|/\mu^{\hspace{0.1mm}2}_{k,1}\ll1$, 
the entire spectrum of Rindler modes
is approximately spanned. For instance, for sample values characteristic of relativistic 
maximally mixed (i.e. $\theta=\pi/4$) particles with squared mass difference  $10^{-3}\,\mathrm{eV}^2$ and typical energy $1\,\mathrm{GeV}$,\footnote{These are typical values for atmospheric neutrinos~\cite{RPDG}.} 
the lower bound on $\Omega_n$ takes the value
$\Omega^{(min)}_n=|\Delta m^2|\sin^2\theta/(2\pi^2\mu^{\hspace{0.1mm}2}_{k,1})\simeq10^{-23}$.
On the other side, the departure from extensive
thermodynamics becomes increasingly negligible
for large $\Omega_n$, restoring the Boltzmann-Gibbs
theory in the limit $\Omega_n\rightarrow\infty$. This is
consistent with the fact that, 
the higher the energy of the state, the lower the average number of particles that can be stored, with both the 
standard and $q$-generalized distributions approaching
zero as $\Omega_n$ increases. 
As a consequence, the difference between the two spectra 
is expected to shrink as $\Omega_n\rightarrow\infty$.

Now, the relation~\eqref{q1} states that $q<1$
within our analysis. To see whether the 
constraint~\eqref{constraint}
is fulfilled, let us employ Eqs.~\eqref{EnRind} and~\eqref{q1}.
A direct substitution in the upper condition yields
\begin{equation}
0\le a\hspace{0.1mm}\Omega_n\le \frac{\pi \mu^{\hspace{0.1mm}2}_{k,1}}{|\Delta m^2|\hspace{0.2mm}\sin^2\theta}\,a\hspace{0.1mm}\Omega_n\hspace{3mm}\Longrightarrow\hspace{3mm}0\le1\le \frac{\pi \mu^{\hspace{0.1mm}2}_{k,1}}{|\Delta m^2|\hspace{0.2mm}\sin^2\theta}\,,
\end{equation}
which is indeed satisfied in the approximation
$|\Delta m^2|/\mu^{\hspace{0.1mm}2}_{k,1}\ll1$.

\begin{figure}[t]
\resizebox{5.3cm}{!}{\includegraphics{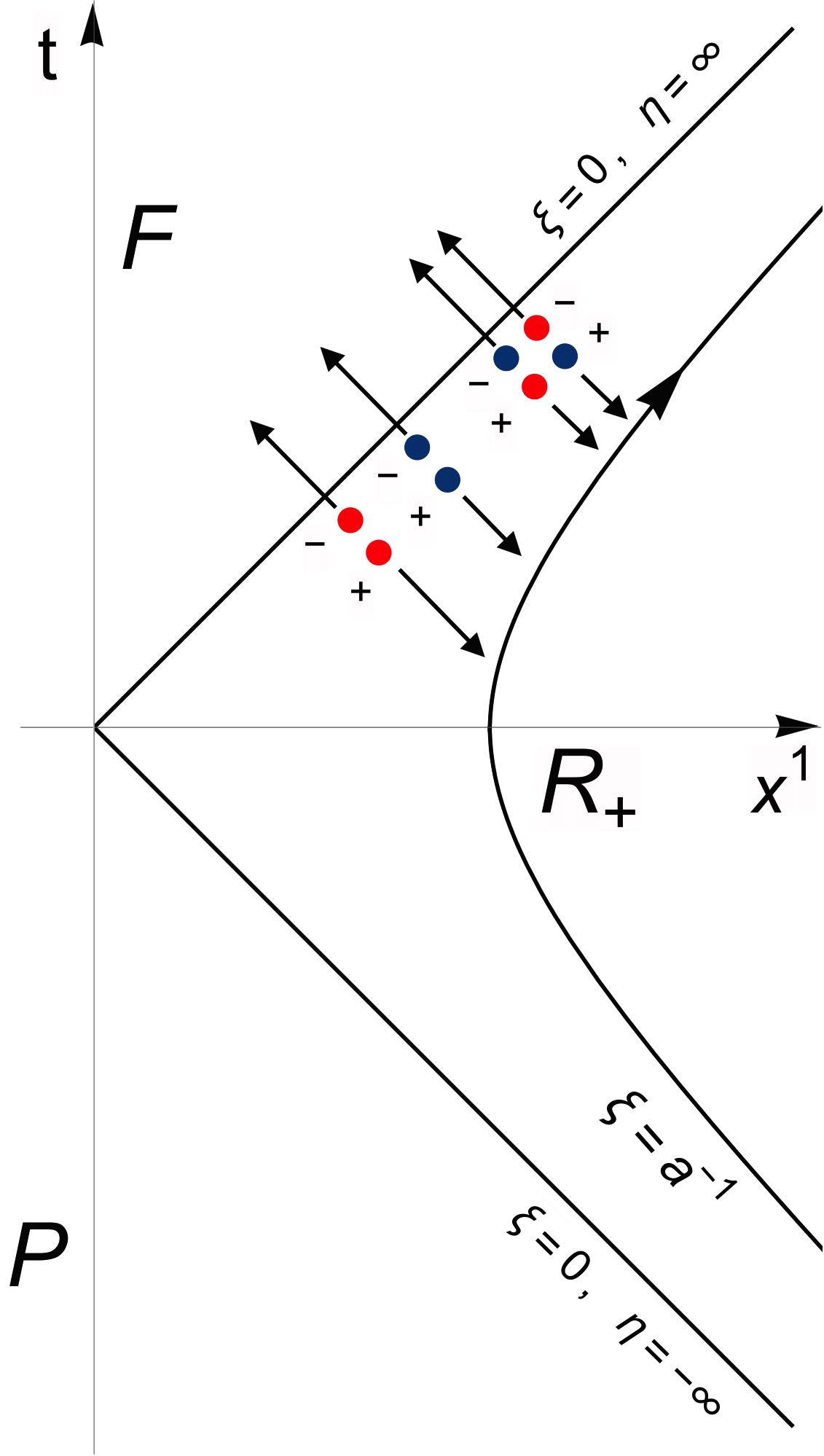}}
\caption{\small{Pictorial interpretation of Unruh spectrum  for mixed fields. Different (online) dot colors correspond to different particle-antiparticle flavors. Unruh effect originates from
quantum vacuum fluctuations close to Rindler horizon.
In the absence of mixing, vacuum contains a single type of particle-antiparticle pairs (either blue-blue or red-red). Conversely, for mixed fields hybrid types of pairs (red-blue and blue-red) do appear as well. This spoils the characteristic particle spectrum, which can be approximately identified with the $q$-generalized Bose-Einstein condensate based on Tsallis entropy.}}
\label{Rindlerbis}
\end{figure}

The connection between 
the perturbed spectrum~\eqref{newmodBE}
and the $q$-generalized Bose-Einstein distribution~\eqref{Loq}
can be explained in terms of the entangled 
structure acquired by the Minkowski vacuum
for mixed fields. As discussed in the Appendix, 
the vacuum for definite flavor fields
becomes a condensate of entangled particle-antiparticle
pairs having both equal and different masses~\cite{BV95,BlasCap}.
Consequently, while
the standard Unruh effect arises from 
one-type fluctuations popping out near the Rindler horizon
(one element of which crossing the horizon, 
the other escaping in the form of Unruh radiation), 
for mixed fields it can be generated by different types of 
entangled pairs (see Fig.~\ref{Rindlerbis}).
This further degree of freedom results into an increase of the
total entropy of the system, which in turn
alters the characteristic number spectrum of particles. 
As shown above, such an effect can be 
described by modeling the new vacuum distribution 
according to Tsallis $q$-thermodynamics
rather than Boltzmann one, 
the departure being proportional to the mixing
parameters and the energy scale (see Eq.~\eqref{q1}). 
From Eq.~\eqref{sqab}, we indeed notice that
having $q<1$ amounts to saying that the 
entropy function associated to the vacuum condensate
of $\phi_A$-quanta (e.g. the blue-blue pairs) and $\phi_B$-quanta 
(the red-red pairs)
is higher than the sum of the entropies 
associated to the condensates of the
two fields separately, due to the presence of
hybrid (blue-red and red-blue) particle-antiparticle pairs.
We stress that this is a peculiar field theoretical effect
boiling down to the nonfactorizability of Fock space for flavor fields,
including the vacuum state~\cite{Cabo} (see also Appendix). 

From the above considerations, we infer that 
the correlations induced by mixing do spoil
the macroscopic properties of Unruh 
thermal condensate by affecting the 
statistical behavior of its microscopic configurations. 
This gives rise to an entangled condensate structure 
of both equal and 
hybrid particle-antiparticle pairs that obey the nonadditive 
Tsallis entropy law in a suitable limit. 
Remarkably, we notice that nonextensive
statistics based on Tsallis entropies have been largely used 
in the study of entanglement~\cite{TSENT},
such as the relative entropy and the 
Peres criterion.

The above result is
quite general, since it is not confined to bosons solely.
In~\cite{NonTN} the Unruh effect has been investigated 
in the case of mixing of (Dirac) fermions, and in particular
of neutrinos, showing that
the vacuum condensate of Rindler particles should be
modified as (see Eq.~(23) of~\cite{NonTN})
\be
\label{modFD}
\mathcal{N}_{\theta,\Delta m}(\Omega_n)\,=\,N_{\mathrm{FD}}(\Omega_n)\,+\,\sin^2\theta\,\,\mathcal{O}\hspace{-0.6mm}\left(\frac{|\Delta m^2|}{\mu^2_{k,1}}\right)\,=\,\frac{1}{e^{2\pi\Omega_n}+1}\,+\,\sin^2\theta\,\,\mathcal{O}\hspace{-0.6mm}\left(\frac{|\Delta m^2|}{\mu^2_{k,1}}\right), 
\ee
where the zeroth-order term is now given by the Fermi-Dirac
distribution function, while the higher-order corrections $\mathcal{O}\left({|\Delta m^2|}/{\mu^2_{k,1}}\right)$
depend on the convolution integral of the condensation density
of mass vacuum in such a way that the Pauli principle is still satisfied. In particular, we have~\cite{NonTN}
\begin{eqnarray}
&&\hspace{2.65cm}\mathcal{O}\left(\frac{|\Delta m^2|}{\mu^2_{k,1}}\right)\,=\,\frac{e^{\pi\Omega_n}}{2\hspace{0.1mm}\cosh\left(\pi\Omega_n\right)}\hspace{0.9mm}N_{F,F}(\Omega_n)\,-\,\frac{e^{-\pi\Omega_n}}{2\hspace{0.1mm}\cosh\left(\pi\Omega_n\right)}\hspace{0.9mm}N_{G,G}(\Omega_n)\,,\\[3mm]
&&N_{F,F}(\Omega_n)=\sum_{r=1,2}\int\,F^{\hspace{0.2mm}*}_{r}(k_1,\Omega_n)\,F_{r}(k_1,\Omega_n)\,{|V_\bk|}^2,\quad\,\, N_{G,G}(\Omega_n)=\sum_{r=1,2}\int\,G^{\hspace{0.2mm}*}_{r}(k_1,\Omega_n)\,G_{r}(k_1,\Omega_n)\,{|V_\bk|}^2.
\end{eqnarray}
Here $F_r$ and $G_r$ are proper combinations
of Dirac modes of spin $r$ in Minkowski and Rindler quantizations~\cite{NonTN}
and 
\be
V_\bk(t)=\left(\frac{\omega_{{\bf k},1}+m_{1}}{2\omega_{{\bf k},1}}\right)^{\frac{1}{2}}
\left(\frac{\omega_{{\bf k},2}+m_{2}}{2\omega_{{\bf k},2}}\right)^{\frac{1}{2}}\left(\frac{|{\bf k} |}{\omega_{{\bf k},2}+m_{2}}-\frac{|{\bf k}|}{\omega_{{\bf k},1}+m_{1}}\right)\,e^{i(\omega_{{\bf k},2}+\omega_{{\bf k},1})\,t},
\label{eqn:Vk2}
\ee
is the analogue of the Bogoliubov coefficient $\bogvb(t)$ in Eq.~\eqref{bogoco}
for fermion mixing~\cite{BV95} (we have assumed for simplicity $\textbf{k}=(k_1,0,0))$. Once more, 
it is easy to check that Eq.~\eqref{modFD} 
reproduces the standard result for $\theta\rightarrow0$
and/or $m_1\rightarrow m_2$, consistently with the absence of mixing in both cases. The same holds in the quantum mechanical 
limit of large momenta with respect to the mass difference, 
since the condensation density $|V_\bk|^2\rightarrow0$.

At the same time, the $q$-modified Fermi-Dirac
distribution in Tsallis thermostatistics can be approximately 
written as~\cite{Buyu3}
\be
\label{modFD}
N_q(\epsilon_n)=\frac{1}{\left[1+(q-1)\beta \epsilon_n\right]^{1/(q-1)}+1}\,.
\ee
Therefore, by expanding for small deviations of $q$ from unity
and following the same reasoning as earlier, we arrive
at a relation akin to Eq.~\eqref{q1} for fermions. 
Clearly, such an extension deserves careful attention, 
since neutrinos are the most abundant and emblematic example
of mixed particles.

\section{Conclusions and Outlook} 
\label{Conc}

The Unruh effect predicts that 
a uniformly accelerated observer 
measures a Planck emission
distribution in Minkowski vacuum. 
However, for quantum fields exhibiting 
entanglement correlations induced by mixing, 
this result turns out to be nontrivially spoilt~\cite{Luciano,NonTN}. 
Here we have discussed this phenomenon 
from a statistical point of view.
Working in the approximation of small mass difference
between the mixed fields, we have shown that 
the modified vacuum distribution 
can be modeled by the $q$-generalized Bose-Einstein
distribution based on the nonadditive Tsallis entropy. In 
this effective description, the deviation from Planckianity 
is found to be quantified by the mixing angle $\theta$ and the mass difference $\Delta m$. Furthermore, 
the $q$-entropic index exhibits 
a running behavior, which is reasonably expected for QFT systems as discussed in~\cite{App13}. The outcome that $q<1$ indicates
that we are in the superadditive regime of Tsallis statistics, 
consistently with the appearance of both equal
and hybrid particle-antiparticle pairs in the vacuum.  

Apart from more formal aspects, we remark that the above picture 
allows us to extend the peculiar 
thermal features of Unruh effect
to mixed fields. Indeed, in~\cite{Luciano,NonTN}
flavor mixing was seen as responsible 
for the emergence of nonthermal contributions in the
Unruh spectrum. Here we have proved that
the origin of these extra terms can be 
explained in terms of a departure of the vacuum 
distribution from Boltzmann-Gibbs
statistics. In turn, this phenomenon is attributable
to  the complex structure acquired by the
vacuum state for mixed fields, which becomes
a condensate of entangled particle-antiparticle pairs
of different species. In other words, we can still identify a temperature 
for the vacuum distribution, provided 
that we work in the framework of Tsallis's thermostatistics.
Nevertheless, following~\cite{AbePla} 
we point out that the new physical temperature
would be different from Unruh temperature by a factor
depending on the nonextensivity $q$-parameter
and the modified entropy $S_q$, in such a way 
that the usual result is still recovered for $q\rightarrow1$.
In this regard, we mention that a similar $q$-dependent
expression for Unruh temperature in Tsallis's theory
has been obtained in~\cite{Shababi} in the context of the Generalized Uncertainty Principle. Possible connections between
the two results need further consideration and
will be addressed elsewhere. 

In passing, we highlight 
that a nonthermal behavior of Unruh effect has been
recently exhibited in~\cite{Dop} even for
the case of a single (i.e. unmixed) massive
field. In that case, it has been found that, 
contrary to what happens with a linear dispersion relation 
characteristic of massless fields, the thermality of Unruh
condensate would be lost for more general 
dispersion relations including a mass term, 
unless one defines  a varying apparent Unruh temperature depending on both the acceleration and the degree of departure from linearity. 
Therefore, it would be interesting to investigate
whether such a result interfaces with 
our reformulation of Unruh effect in Tsallis's
language. For this purpose, however, 
a formalism based on the relativistic Doppler shift
method is required~\cite{Dop}, since the Bogoliubov transformation approach is insensitive to 
the mass of the field when computing 
Unruh vacuum distribution~\cite{Takagi}.

Beyond the above issues, several other aspects
remain to be analyzed. To avoid unnecessary technicalities, 
we have focused on a simplified model involving only two scalar
fields, noticing that similar considerations can be extended
to fermions quite straightforwardly.
Furthermore, our perturbative 
analysis relies on the leading-order approximation of 
small difference between the masses of mixed fields. 
The question thus arises as to how the connection~\eqref{q1}
between the nonextensive $q$-index and the mixing
parameters would appear for arbitrary mass differences,
as well as in the case of three flavor generations. 
Another extension is to apply the
above formalism to the best-known
Hawking radiation, which has
been largely studied within the framework of 
nonextensive corrected-entropies
in recent years~\cite{Barrow}. 

From a more phenomenological perspective, 
it would be challenging to test possible experimental
implications of our result. As well known, direct evidences
of Unruh effect have not yet been obtained, the obvious reason
being the fact that the Unruh temperature is extremely small
even for huge accelerations. 
However, there have been many proposals
in the literature to bypass hindrances
arising from technical limitations
by focusing on analogues of Unruh effect, 
even at the classical level. For instance, 
feasible tests are being analyzed by simulating
vacuum fluctuations of Minkowski spacetime 
through gravity waves on
the surface of water subject to white noise~\cite{Leona}.
Attempts to detect indirect traces of Unruh radiation have also been
carried out in graphene~\cite{graphene} and
metamaterials~\cite{Smoly}, where the effects of 
Rindler-like horizons are mimicked by
means of photons waveguides. Thus, such analog models 
provide the only test bench for probing the Unruh effect and
any possible deviation from the standard behavior to date.

Finally, one more direction to explore is 
whether Tsallis statistics and mixed particles 
are intertwined on a more fundamental level that 
goes beyond the specific framework of Unruh effect. 
In this vein, we emphasize the recent proposal 
to solve the long-standing problem of abundance of primordial ${}^{7}\mathrm{Li}$, 
which is affected by the neutrino interactions and primordial magnetic field, by investigating the impact of Big Bang nucleosynthesis predictions of adopting a Tsallis distribution for the nucleon energies~\cite{Litium}. 
Work along the above research lines
is presently under active consideration. 


\acknowledgements 

One of the authors (GGL) is grateful to Costantino
Tsallis (Centro Brasileiro de Pesquisas Fisicas, Brazil) and Gaetano Lambiase (Universit\`a degli Studi di Salerno, Italy) for helpful conversations.

\appendix
\section{QFT of flavor mixing in Minkowski spacetime}
\label{Qmix}
We review the QFT formalism of flavor mixing 
for the simplest case of two scalar fields
in Minkowski background~\cite{BlasCap}. 
Toward this end, we introduce the algebraic 
generator of mixing
\begin{equation}
G_\theta(t)=\exp\left[-i\theta\int d^3x\hspace{0.2mm}\Big(\pi_1(x)\phi_2(x)\,-\,\pi_2(x)\phi_1(x)\Big) \right],
\label{genmix}
\end{equation}
where the notation has already been 
set up in Sec.~\ref{QFTRIND}. 
In terms of this operator, the mixing transformations in Eqs.~\eqref{Ponteca} and \eqref{Pontecb} can be cast as
\be
\phi_\ell(x)=G_\theta^{-1}(t)\phi_j(x)G_\theta(t)\,,
\label{mixing2}
\ee
where $(\ell,j)=\left\{(A,1),(B,2)\right\}$. One can prove
that $G_\theta(t)$ belongs to $SU(2)$ group, the algebra of which 
is closed by the operators~\cite{BlasCap}
\begin{equation}
S_+(t)=-i\int d^3x\hspace{0.7mm}\pi_1(x)\phi_2(x)\,,\quad S_{-}(t)=-i\int d^3x\hspace{0.7mm}\pi_2(x)\phi_1(x)\,,
\end{equation}
\vspace{-4mm}
\begin{equation}
S_3=-\frac{i}{2}\int d^3x\hspace{0.7mm}\Big(\pi_1(x)\phi_1(x)-\pi_2(x)\phi_2(x)\Big)\,,\quad S_0=-\frac{i}{2}\int d^3x\hspace{0.7mm}\Big(\pi_1(x)\phi_1(x)+\pi_2(x)\phi_2(x)\Big)\,.
\end{equation}

Within the framework of QFT mixing, the generator~\eqref{genmix}
plays a pivotal r\^ole, 
as it provides the dynamical map between
the Fock space $\mathcal{H}_{A,B}$ for the fields 
with definite flavor and the Fock space
$\mathcal{H}_{1,2}$ for the fields with definite mass\footnote{For brevity, henceforth $\mathcal{H}_{A,B}$ and $\mathcal{H}_{1,2}$ are simply referred to as ``flavor'' and ``mass'' Fock space, respectively.}.
Indeed, let us consider 
the generic matrix element of $\mathcal{H}_{1,2}\,=\mathcal{H}_1\otimes\mathcal{H}_2$, i.e. $_{1,2}\langle a|\phi_{i}(x)|b\rangle_{1,2}$ ($i=1,2$), 
where $|a \rangle_{1,2}$ and $|b \rangle_{1,2}$  
are arbitrary states in ${\mathcal H}_{1,2}$. 
By inverting Eq.~(\ref{mixing2}) with respect to $\phi_j$, 
we get
\begin{equation}
_{1,2}\langle a|G_{\theta}(t)\hspace{0.2mm}
\phi_\ell(x)\hspace{0.2mm}G^{-1}_{\theta}(t)|b \rangle_{1,2}=\hspace{0.1mm}_{1,2}\langle a|\phi_{j}(x)|b \rangle_{1,2}\,,
\label{eqn:inversequation}
\end{equation}
which in fact shows that $G_{\theta}^{-1}(t)|b \rangle_{1,2}$ 
is a vector in $\mathcal{H}_{A,B}$. 
Therefore, we can write
\begin{equation}
G^{-1}_{\theta}(t): {\mathcal H}_{1,2} \mapsto {\mathcal H}_{A,B}\,.
\label{mapp}
\end{equation}
In particular, for the vacuum state
$|0_\mathrm{M}\rangle_{1,2}\equiv|0\rangle_{1,2}=|0\rangle_{1}\otimes|0\rangle_{2}$, this gives
\be
|0(\theta,t) \rangle_{A,B} = G^{-1}_{\theta}(t)|0 \rangle_{1,2}\,.  
\label{flavvacu}
\ee
The above relation allows us to define
the time-dependent flavor vacuum  
$|0(\theta,t) \rangle_{A,B}$ 
in terms of the corresponding mass vacuum $|0\rangle_{1,2}$. 

A comment is in order here. For quantum mechanical systems 
(i.e. systems with finite number of degrees of freedom),  
$G_{\theta}(t)$ is a unitary operator that preserves 
the canonical commutation relations. This is ensured
by Stone-von Neuman theorem~\cite{SvN,Stone}, 
which states that any two irreducible representations of the canonical commutators are unitarily equivalent in Quantum Mechanics.
Accordingly, mass and flavor representations
give rise to physically equivalent
descriptions of mixing. On the other hand,  
in QFT the transformation~\eqref{genmix} 
is found to be nonunitary in the infinite volume limit, which
means that the vacua $|0\rangle_{1,2}$ 
and $|0(t)\rangle_{A,B}$ become mutually
orthogonal and the related Fock spaces
unitarily inequivalent. This is quite
different from the conventional perturbation theory, 
where the vacuum of the interacting theory is
expected to be essentially the same as that of 
the free theory (up to a phase
factor)~\cite{BogoCit}. Clearly, such an inequivalence and its implications
disappear for $\theta=0$ and/or $m_2=m_1$, consistently with
the fact that there is no mixing in both cases.

To find out how the mapping~\eqref{flavvacu}
affects the structure of the flavor
vacuum, we now focus on the derivation of ladder operators in the flavor basis. By using the standard plane-wave
quantization~\eqref{eqn:expans0} for both $\phi_1$ and $\phi_2$,  
Eq.~\eqref{mixing2} leads to the following
expansions for the flavor fields
\be
\label{eqn:dynamicalmap2}
\phi_\ell(x)=\int d^{3}{k}\hspace{0.2mm}\Big\{a_{\textbf{k},\ell}(\theta, t)\, \uuu_{\textbf{k},j}(x)+ \bar a_{\textbf{k},\ell}^\dagger(\theta, t)\, \uuu_{\textbf{k},j}^{\hspace{0.1mm}*}(x) \Big\},\,\, \quad (\ell,j)=\left\{(A,1),(B,2)\right\},
\ee
where
\be
\label{tra}
a_{\textbf{k},\ell}(\theta, t)\,\equiv\, G^{-1}_{\theta}(t)\hspace{0.2mm}a_{\textbf k,j}\hspace{0.2mm}G_{\theta}(t)\,, 
\ee
is the annihilator of a quantum with definite flavor $\ell$
(for simplicity, we refer to this operator as flavor annihilator
and use the handier notation $a_{\textbf{k},\ell}(\theta, t)\equiv a_{\textbf{k},\ell}(t)$). From the above relation, 
we obtain 
\be
a _{{\bk},A}(t)=\cos\theta\, a _{{\bk},1}\,+\,\sin\theta\, \left(\bogubst(t)\, a _{{\bk},2}\, + \, \bogvb(t)\, \bar a^\dagger _{-{\bk},2}\right),
\label{aka}
\ee
(similarly for $a_{\textbf{k},B}(\theta, t))$. Therefore, 
the flavor annihilator is
related to the corresponding ladder operators
in the mass basis via of a Bogoliubov
transformation (the terms in the brackets) nested into a rotation.
The Bogoliubov coefficients are defined as
\begin{equation}
\label{bogoco}
\bogub(t)=|\bogub|\, \ebogomeno\,,\quad
\bogvb(t)=|\bogvb|\, \ebogopiu\,,
\end{equation}
where 
\be
\label{bocoeffi}
|\bogub|\equiv\frac{1}{2}\left( \sqrt{\frac{\omega_{\bk,1}}{\omega_{\bk,2}}}\,+\,\sqrt{\frac{\omega_{\bk,2}}{\omega_{\bk,1}}}\right),\qquad |\bogvb|\equiv\frac{1}{2}\left( \sqrt{\frac{\omega_{\bk,1}}{\omega_{\bk,2}}}\,-\,\sqrt{\frac{\omega_{\bk,2}}{\omega_{\bk,1}}}\right). 
\ee
It is easy to verify that
\be
{|\bogub|}^2\,-\,{|\bogvb|}^2=1\,, 
\label{eqn:iper}
\ee
which ensures that the flavor operator~\eqref{tra}
and its conjugate are still canonical (at equal times).

The mapping~\eqref{flavvacu}
induces a physically nontrivial structure in the
flavor vacuum, which becomes an entangled $SU(2)$ 
coherent state made up by
particle-antiparticle pairs both of the same
and different masses~\cite{BV95}.  In turn, 
this inequivalence affects the well-known oscillation formula to include the antiparticle degrees of freedom~\cite{BlasPlb}. 
The condensation density of flavor vacuum
is given by
\begin{equation}
\label{eqn:denscondbosoni2} 
{}_{A,B}\langle 0(t)| a_{{\bk'},i}^{\dagger}\, a_{{\bf
k},j} |0(t)\rangle_{A,B}= \sin^{2}\theta\; {|\bogvb|}^2\delta_{ij}\delta^3({\textbf{k}-\textbf{k}'})\,.
\end{equation}
Clearly, by exploiting the symmetric structure of
Eq.~\eqref{tra}, one can reverse the above reasoning
and analyze the properties of mass
vacuum, which appears as a condensate of
particle-antiparticle pairs having both equal and different flavors.
In line with our previous considerations 
on the disappearance of the inequivalence
for vanishing mixing, the condensation density~\eqref{eqn:denscondbosoni2}  
goes to zero for $\theta=0$ (since $a_{\bk,A}\rightarrow a_{\bk,1}$ and 
$a_{\bk,B}\rightarrow a_{\bk,2}$)
and/or $m_2=m_1$ (since the Bogoliubov
coefficients reduce to
$\bogub(t)=1$ and  $\bogvb(t)=0$, which in turn implies
that $a_{\bk,A}$ and $a_{\bk,B}$ are simple
superpositions of $a_{\bk,1}$ and $a_{\bk,2}$). 
Notice that the same behavior
occurs for ${|\bk|}^2\gg\frac{{m_1}^2+{m_2}^2}{2}$, 
thus allowing to recover the standard quantum mechanical 
description of mixing
in the relativistic approximation.

The complex structure of the flavor vacuum
$|0(t)\rangle_{A,B}$ has been recently 
studied in~\cite{Cabo}, 
where it has been
established that the Fock space for flavor fields cannot be obtained by the direct product of the spaces for massive
fields. This strengthen the result that entanglement properties
for mixed fields already emerge
at the level of vacuum state. As a remark, the observation that flavor 
mixing can be associated with (single-particle) entanglement 
traces back to~\cite{Dimauro} and has inspired 
a series of studies on violations of
Bell, Leggett-Garg and Mermin-Svetchlichny inequalities, nonlocality, gravity-acceleration degradation effects and
other similar phenomena~\cite{SimPhe1,SimPhe2,SimPhe3,SimPhe4,SimPhe5}.
The entanglement content of the flavor vacuum 
has been explicitly quantified in~\cite{Vacent} in the 
limit of small mass difference
and/or mixing angle.

\end{document}